\documentclass[fleqn,twoside]{article}
\usepackage{espcrc2}
\usepackage{graphics}
\input{psfig.sty}

\usepackage{graphicx}
\usepackage[figuresright]{rotating}

\newcommand{\AmS}{{\protect\the\textfont2
  A\kern-.1667em\lower.5ex\hbox{M}\kern-.125emS}}

\def\simlt{\mathrel{\lower2.5pt\vbox{\lineskip=0pt\baselineskip=0pt
           \hbox{$<$}\hbox{$\sim$}}}}
\def\simgt{\mathrel{\lower2.5pt\vbox{\lineskip=0pt\baselineskip=0pt
           \hbox{$>$}\hbox{$\sim$}}}}
\newcommand{\Tr}{\hbox{\rm Tr}}
\def\simge{\mathrel{%
   \rlap{\raise 0.511ex \hbox{$>$}}{\lower 0.511ex \hbox{$\sim$}}}}
\def\simle{\mathrel{
   \rlap{\raise 0.511ex \hbox{$<$}}{\lower 0.511ex \hbox{$\sim$}}}}
 
\def\slashchar#1{\setbox0=\hbox{$#1$}           
   \dimen0=\wd0                                 
   \setbox1=\hbox{/} \dimen1=\wd1               
   \ifdim\dimen0>\dimen1                        
      \rlap{\hbox to \dimen0{\hfil/\hfil}}      
      #1                                        
   \else                                        
      \rlap{\hbox to \dimen1{\hfil$#1$\hfil}}   
      /                                         
   \fi}                                         %
\def\nn{\nonumber}
\def\ts{\thinspace}

\def\ra{\rightarrow}

\def\be{\begin{equation}} 
\def\ee{\end{equation}} 
\def\bea{\begin{eqnarray}}
\def\eea{\end{eqnarray}}
\def\ba{\begin{array}}
\def\ea{\end{array}}
\def\dag{\dagger}

\def\CH{{\cal H}}

\def\CM{{\cal M}}

\def\CO{{\cal O}}

\def\CW{{\cal W}}

\def\kslash{\raise.15ex\hbox{/}\kern-.57em k}

\def\tev{{\rm TeV}}

\def\half{{\textstyle{ { 1\over { 2 } }}}}

\title{
\vskip -15mm
\begin{flushright}
\vskip -15mm
{\small BUHEP-02-30\\
hep-ph/0210240\\}
\vskip 5mm
\end{flushright}
Deconstructing Dimensional Deconstruction}

\author{Kenneth Lane \\
{Department of Physics\\
Boston University\\
590 Commonwealth Avenue\\
Boston, MA 02215, USA}%
\thanks{Email: lane@bu.edu}}
       
\begin{document}

\begin{abstract}
Dimensional deconstruction (DD) abstracts from higher dimensional models
features of related 4--dimensional ones. DD was proposed in
Refs.~\cite{acga,acgb,hilletal} as a scheme for constructing models of {\it
naturally} light composite Higgs boson. These are models in which---{\it
without fine--tuning of parameters}---the composite Higgs's mass $M$ and
vacuum expectation value $v$ are much lighter than its binding energy scale
$\Lambda$. We review the basic idea of DD. It is easy to arrange $M \ll
\Lambda$. We show, however, that DD fails to give $v \ll \Lambda$ in a model
that is supposed to contain a naturally light composite Higgs~\cite{csdd}.
\vspace{1pc}
\end{abstract}

\maketitle

\section{WHAT IS DIMENSIONAL DECONSTRUCTION?}

There has been considerable interest lately in a new approach to
model--building called ``dimensional deconstruction'' (DD). In the beginning,
there were two views of DD. The one we discuss in this paper is due to
Arkani-Hamed, Cohen and Georgi (ACG)~\cite{acga,acgb}. It is based on the
fact that certain renormalizable, asymptotically free 4d field theories look,
{\it for a limited range of energies}, like $d > 4$--dimensional theories in
which the extra dimensions are compactified and discretized (on a periodic
lattice). Here, the extra dimensions are a mirage. The other view is that of
Hill and his collaborators~\cite{hilletal} who assume the extra dimensions
are real. They discretize the extra dimensions too---to regulate the
theory. Both Arkani-Hamed {\it et al.} and Hill {\it et al.} use features of
the higher dimensional model to deduce the form, magnitude, and sensitivity
to high--scale ($\Lambda$) physics of phenomenologically important operators
such as mass terms (generically, $M$), self--interactions ($\lambda$), and
vevs ($v$) of light {\it composite} Higgs bosons (LCH)~\cite{lch}.  Their LCH
models aim for $M\simeq v \simeq 100$--200~GeV and $\Lambda \simeq 10\,\tev$,
relevant to electroweak symmetry breaking.

\begin{figure}[tb]
\vbox to 7.5cm{
\vfill
\includegraphics{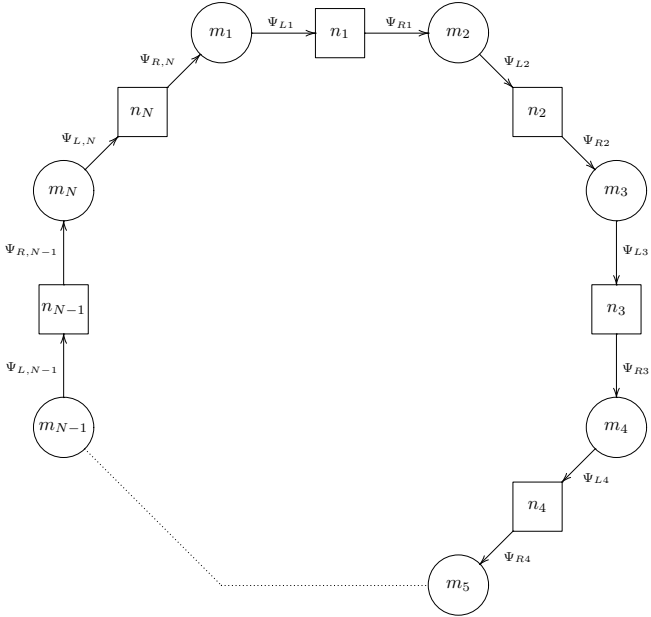}
\vfill
}
\caption{The full moose for the ring model of Ref.~\cite{acga}, showing its
  UV completion. Strong gauge groups are labeled by $n_1,\dots,n_N$ and weak
  gauge groups by $m_1,\dots,m_N$.
\label{fig:a}}
\end{figure}

The simplest DD example is the $d=5$ ``moose ring'' model~\cite{acga}
depicted in Fig.~1. This shows the full content (``UV--completed'') of the
model at the high energy scale $\Lambda$. It contains $N$ strong $SU(n)$ and
weak $SU(m)$ (coupling $g^2/4\pi\ll 1$) gauge groups, with matter fields that
are the massless chiral fermions
\bea\label{eq:fermionsa}
&& \psi_{Lk} \in (n,m,1)\ts, \ts\ts\ts \psi_{Rk} \in (n,1,m) \nn \\
&&\qquad {\rm of} \ts\ts\ts (SU(n)_k, \ts SU(m)_k, \ts SU(m)_{k+1}) \ts.
\eea
The index $k$ is periodically identified with $k+N$. As $g\ra 0$, these
fermions have a large chiral symmetry, $[SU(m)_L \otimes SU(m)_R]^N$. At
$\Lambda$, the strong $SU(n)$ interactions cause them to condense, creating
$N$ sets of $m^2-1$ {\it composite} Goldstone bosons (GBs), $\pi_k^a$ with
$k=1,\dots,N$ and $a=1,\dots,m^2-1$. Their decay constant $f \simeq
\Lambda/4\pi$.

Below $\Lambda$, this is a nonlinear sigma model, with fields $U_k =
\exp{(i\pi_k^a t_a/f)} \equiv \exp{(i\pi_k/f)}$ interacting with the
weakly--coupled $SU(m)_k$ gauge fields $A_{k\mu} = A_{k\mu}^a t_a$. They
transform as $U_k \ra W_k U_k W^\dag_{k+1}$ with $W_k \in SU(m)_k$. This low
energy theory is represented by the ``condensed moose'' obtained from Fig.~1
by erasing the $SU(n)$ squares and linking the $SU(m)_k$ and $SU(m)_{k+1}$
circles by $U_k$.

Now, $N-1$ gauge boson multiplets eat $N-1$ sets of GBs and acquire the
masses $\CM_k = 2gf \sin(k \pi/N)$ for $k=1,\dots,N$. The massless gauge
field $A^a_\mu = (A^a_{1\mu} + \cdots + A^a_{N\mu})/\sqrt{N}$ couples with
strength $g/\sqrt{N}$ and the uneaten GB is $\pi^a= (\pi^a_1 + \cdots +
\pi^a_N)/\sqrt{N}$. In the unitary gauge, the 4d theory below $\Lambda$ is
described by uniform link variables $U_k =\exp{(i\pi^a t_a/\sqrt{N}f)}$ plus
the massless and massive gauge fields.

Alternatively, at energies well below $gf$, this looks like a 5d gauge
theory: The fifth dimension is compactified on a discretized circle,
represented exactly by the {\it condensed} moose. For $k\ll N$, there is a
Kaluza--Klein tower of gauge excitations with masses $\CM_k = 2\pi
gfk/N$~\cite{acga}. The circumference of the circle is $R=Na$ where the
lattice spacing $a = 1/gf$ and the 5--dimensional gauge coupling is $g_5^2 =
g^2 a$. The fifth component of the gauge boson $A_5^a = g\pi^a/\sqrt{N}$. The
geometrical connection is clear: $\pi^a$ is the zero mode associated with
rotation about the circle of $SU(m)$ groups in four dimensions and it
corresponds to the fifth--dimensional gauge freedom associated with
$A_5^a$. At higher energies, $\sim f$ or $\Lambda$, the fifth dimension is
deconstructed as the underlying asymptotically free 4d theory appears.

\section{WHAT IS DD GOOD FOR?}

But, $\pi^a$ is really a 4d pseudoGoldstone boson (PGB) whose symmetry is
explicitly broken by the weak $SU(m)_k$ interactions. So it might be a
candidate for the LCH of electroweak symmetry breaking. To be a truly {\it
natural} LCH, its vev $v \ll \Lambda$ also. This requires its quartic
couplings $\lambda \simeq M^2/v^2 = \CO(1)$ or, at least,
not $\ll 1$. The idea of DD is that the magnitude and $\Lambda$--dependence
of $M^2$ and $\lambda$ can be deduced from the higher dimensional
theory. Let's see.

Higher dimensional gauge invariance allows mass for $A_5$ from $|\CW|^2$,
where $\CW = P\exp{(i\int dx_5 A_5)}$ is the nontrivial Wilson loop around
the fifth dimension~\cite{acgb}. Since $|\CW|^2$ is a nonlocal operator, it
cannot be generated with a UV--divergent coefficient. On the discretized
circle, $\CW = \Tr [\Pi_{k=1}^N \exp{(ia A_{5k})}]$. In the 4d theory this is
just the gauge--invariant ${\Tr}(U_1 U_2\cdots U_N)$, and so this is what
provides the mass for $\pi^a$.  Standard power counting indicates that the
strength of $|{\Tr}(U_1 U_2\cdots U_N)|^2$ is $\Lambda^2 f^2
(g^2/16\pi^2)^N$. This is correct only for $N=1$. For $N \ge 2$ infrared
singularities from the gauge boson masses at $g \ra 0$ overcome this power
counting. For $N=2$, $M^2 \sim g^4 f^2 \log(\Lambda^2/\CM^2_B) \sim g^4 f^2
\log(N^2/g^2)$ where $\CM^2_B \sim g^2 f^2 /N^2$ is a typical $SU(m)$ gauge
boson mass. For $N \ge 3$, $M^2 \sim g^4 f^2$. Thus, for $N \ge 2$ and
$g^2/4\pi \sim 10^{-2}$, we have $M \ll \Lambda$, as desired.

DD predicts that $\pi^a$ will fail as an LCH because the quartic interactions
of $A_5$ are derivatively coupled and/or induced by weak $SU(m)$
interactions. This is true for $\pi^a$ as well. Since $p/f \sim M/f \sim
g^2$, all quartic couplings of $\pi^a$ are $\le \CO(g^4)$. So, in this model,
DD is a reliable guide. To achieve larger $\lambda$, ACG applied DD to a 6d
model with nonderivative PGB interactions~\cite{acgb}.

\begin{figure}[tb]
\vbox to 7.0cm{
\vfill
\includegraphics{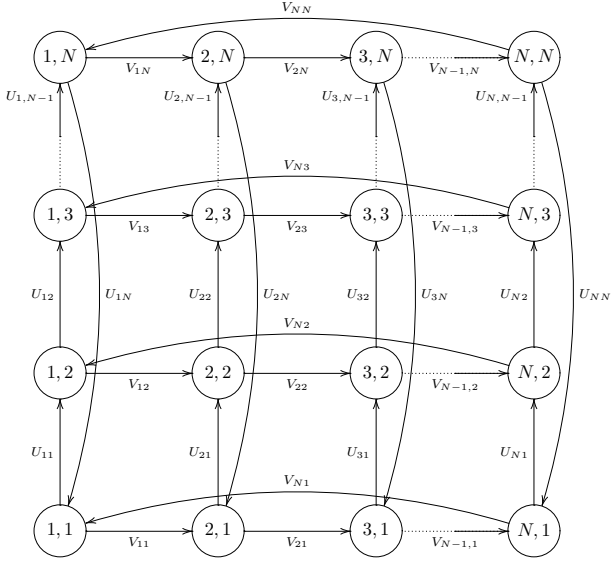}
\vfill
}
\caption{The condensed moose for the 6d toroidal model of Ref.~\cite{acgb}.
\label{fig:b}}
\end{figure}

\section{THE $6d$ TOROIDAL MOOSE MODEL.}

Consider a 4d theory described below its UV--completion scale $\Lambda$ by
the condensed moose in Fig.~2. This resembles a discretized torus with
$N\times N$ sites labeled periodically by integers $(k,l)$. Weakly--coupled
($g$) gauge groups $SU(m)_{kl}$ at the sites are linked by nonlinear sigma
model fields $U_{kl}$ and $V_{kl}$, transforming as
\bea\label{eq:UVtransform}
&&U_{kl} = \exp{(i\pi_{u,kl}/f)} \ra W_{kl} \ts U_{kl} \ts W^\dag_{k,l+1}\ts,
\nn\\
&&V_{kl}  = \exp{(i\pi_{v,kl}/f)} \ra W_{kl} \ts V_{kl} \ts W^\dag_{k+1,l}
\ts.
\eea
The $\pi_{u,kl}$ and $\pi_{v,kl}$ comprise $2N^2$ $SU(m)$ adjoints of
composite Goldstone bosons.

The $SU(m)_{kl}$ gauge bosons eat $N^2-1$ sets of GBs. The spectrum of
massive gauge bosons, $\CM^2_{kl}= 4g^2f^2[\sin^2(k\pi/N) + \sin^2(l\pi/N)]$
is KK--like for small $k,l$. The massless gauge boson is $B^\mu_{NN} = N^{-1}
\sum_{k,l} A^\mu_{kl}$ and its coupling is $g/N$. Among the $N^2+1$ leftover
PBS, two that ACG proposed as light composite Higgses are
\be\label{eq:piupiv}
\pi_u = {1\over{N}}\sum_{k,l} \pi_{u,kl} \ts, \qquad
\pi_v = {1\over{N}}\sum_{k,l} \pi_{v,kl} \ts.
\ee
These are the zero modes associated with going around the torus in the $U$
and $V$--directions.

What does DD predict for the masses and couplings of $\pi_{u,v}$? Viewing the
condensed moose as the compactified and discretized dimensions 5,6 of a 6d
gauge theory, the extra--dimensional gauge fields are $A^a_{5,6} =
g\pi^a_{u,v}/N$. As before, DD predicts small $M^2_{\pi_{u,v}} \sim g^4 f^2
\log(N^2/g^2)$ for $N=2$ and $g^4 f^2$ for $N \ge 3$.

In the 6d model, $A_{5,6}$ have moderately strong nonderivative
interactions~\cite{acgb}. They come from the term $\Tr F_{56}^2 =
\Tr([A_5,A_6]^2) + \cdots = \lambda \Tr([\pi_u,\pi_v]^2) + \cdots$ which, in
turn, arises from the ``plaquette'' Hamiltonian
\be\label{eq:plaq}
\CH_{\rm P} = \sum_{k,l} \lambda_{kl} \ts f^4 \ts \Tr\left(U_{kl} V_{k,l+1}
U_{k+1,l}^\dag V_{kl}^\dag\right) + {\rm h.c.}
\ee
Note that $\CH_{\rm P}$ leaves $\pi_{u,v}$ massless.

In 6d, the quartic coupling may be shown to be $\lambda \equiv
\half\sum_{k,l} \lambda_{kl}/N^4 = g^2/2N^2$~\cite{csdd}. Depending on the
$N$--dependence of the Higgs masses, this may be large enough to give a Higgs
vev comparable to $M_{\pi_{u,v}}$. In 4d, this prediction of DD fails. The
strength of $\lambda$ depends entirely on the nature of the toroidal moose
model's UV completion.

The most natural UV completion of this model is the analog of Fig.~1: At
$\Lambda$, there are $2N^2$ massless fermions $\psi$ with strong $SU(n)$
interactions located midway between the weak $SU(m)$'s~\cite{csdd}. Then, the
plaquette interaction arises only from weak gauge interactions. It is of
$\CO(g^4)$ and, so, $v^2 \sim M_{\pi_{u,v}}^2/\lambda \gg M_{\pi_{u,v}}^2$.

It is possible to find UV completions of the toroidal moose that yield larger
$\lambda$. They involve elementary scalars and, therefore, supersymmetry to
avoid unnatural fine--tuning of parameters~\cite{csdd,agchmg}. More
generally, one can construct sigma models whose symmetries are tailored to
give an effective Lagrangian with {\it arbitrarily and separately tunable}
$M^2$ and $\lambda$---at least at the one--loop level. This is the basis of
an interesting new direction that has evolved from
DD~\cite{littleH,schmaltz}. But this approach, called ``little Higgs'', has
nothing to do with the original idea of deconstruction---that the strengths
of a composite Higgs' mass and interactions may be deduced from corresponding
terms in higher dimensional gauge theories. Finding a truly dynamical,
natural way of UV--completing little Higgs models remains one of the greatest
challenges to this new idea for electroweak symmetry breaking.

\section*{Acknowledgements}

This is the written version of my talk at The 31st International Conference
on High Energy Physics, Amsterdam, The Netherlands, July 24--31, 2002. I
thank the organizers for their kind help and solicitude. I have benefitted
greatly from discussions with Nima Arkani-Hamed, Bill Bardeen, Sekhar
Chivukula, Andy Cohen, Estia Eichten, Howard Georgi and Chris Hill. I thank
Gerard 't Hooft and the theory group at the University of Utrecht for their
hospitality and lively discussion. I am grateful to Fermilab and its Theory
Group for a 2001--2002 Frontier Fellowship which supported this research. It
was also supported in part by the U.S.~Department of Energy under
Grant~No.~DE--FG02--91ER40676.


\begin{thebibliography}{9}
%
%
\bibitem{acga} N.~Arkani-Hamed, A.~G.~Cohen and H.~Georgi,
  Phys.~Rev.~Lett.~{\bf 86}, 4757 (2001); hep-ph/0104005.
%
%
\bibitem{acgb} N.~Arkani-Hamed, A.~G.~Cohen and H.~Georgi, Phys.~Lett.~{\bf
B513}, 232 (2001) and hep-ph/0105239~v4; N.~Arkani-Hamed, {\it et al.},
JHEP~0208:020 (2002); hep-ph/0202089.
%
%
\bibitem{hilletal} C.~T.~Hill, S.~Pokorski and J.~Wang,
Phys.~Rev.~{\bf D64}, 105005 (2001); hep-th/0104035; 
H.~C.~Cheng, C.~T.~Hill, S.~Pokorski and J.~Wang,
Phys.~Rev.~{\bf D64}, 065007 (2001); hep-th/0104179; 
H.~C.~Cheng, C.~T.~Hill and J.~Wang,
Phys.~Rev.~{\bf D64}, 095003 (2001); hep-ph/0105323.
%
%
\bibitem{csdd} This talk is largely based on my paper, K.~Lane, {\it A Case
    Study in Dimensional Deconstruction}, Phys.~Rev.~{\bf D65}, 115001
    (2002); hep-ph/0202093.
%
%
\bibitem{lch} D.~B.~Kaplan and H.~Georgi, Phys.~Lett. {\bf B136}, 183 (1984); 
S.~Dimopoulos et al., Phys.~Lett.~{\bf B136}, 187 (1984).
%
%
\bibitem{agchmg} A.~Cohen, H.~Georgi, communications.
%
%
\bibitem{littleH} N.~Arkani-Hamed, {\it et al.}, JHEP~0208:021 (2002),
  hep-ph/0206020 and JHEP~0207:034 (2002), hep-ph/0206021.
%
%
\bibitem{schmaltz} M.~Schmaltz, {\it Beyond the Standard Model (theory)},
  invited talk at the 31st International Conference on High Energy Physics,
  Amsterdam, The Netherlands, July 24-31, 2002.
\end{thebibliography}
\end{document}